

\def\singlespace{\normalbaselines}
\def\oneandahalfspace{\baselineskip=1.15\normalbaselineskip plus 1pt
\lineskip=2pt\lineskiplimit=1pt}

\def\np{\vfill\eject}

\def\nofirstpagenoten{\nopagenumbers\footline={\ifnum\pageno>1\tenrm
\hss\folio\hss\fi}}
\def\nofirstpagenotwelve{\nopagenumbers\footline={\ifnum\pageno>1\twelverm
\hss\folio\hss\fi}}
\def\leaderfill{\leaders\hbox to 1em{\hss.\hss}\hfill}
\def\ft#1#2{{\textstyle{{#1}\over{#2}}}}
\def\frac#1/#2{\leavevmode\kern.1em
\raise.5ex\hbox{\the\scriptfont0 #1}\kern-.1em/\kern-.15em
\lower.25ex\hbox{\the\scriptfont0 #2}}
\def\sfrac#1/#2{\leavevmode\kern.1em
\raise.5ex\hbox{\the\scriptscriptfont0 #1}\kern-.1em/\kern-.15em
\lower.25ex\hbox{\the\scriptscriptfont0 #2}}


\parindent=20pt
\def\narrow{\advance\leftskip by 40pt \advance\rightskip by 40pt}

\def\AB{\bigskip
        \centerline{\bf ABSTRACT}\medskip\narrow}
\def\nonarrower{\advance\leftskip by -40pt\advance\rightskip by -40pt}
\def\AE{\bigskip\nonarrower}

\def\boxit#1{\vbox{\hrule\hbox{\vrule\kern3pt
        \vbox{\kern3pt#1\kern3pt}\kern3pt\vrule}\hrule}}

\def\gtorder{\mathrel{\raise.3ex\hbox{$>$}\mkern-14mu
             \lower0.6ex\hbox{$\sim$}}}
\def\ltorder{\mathrel{\raise.3ex\hbox{$<$}|mkern-14mu
             \lower0.6ex\hbox{\sim$}}}
\def\dalemb#1#2{{\vbox{\hrule height .#2pt
        \hbox{\vrule width.#2pt height#1pt \kern#1pt
                \vrule width.#2pt}
        \hrule height.#2pt}}}

\font\fourteentt=cmtt10 scaled \magstep2
\font\fourteenbf=cmbx12 scaled \magstep1
\font\fourteenrm=cmr12 scaled \magstep1
\font\fourteeni=cmmi12 scaled \magstep1
\font\fourteenss=cmss12 scaled \magstep1
\font\fourteensy=cmsy10 scaled \magstep2
\font\fourteensl=cmsl12 scaled \magstep1
\font\fourteenex=cmex10 scaled \magstep2
\font\fourteenit=cmti12 scaled \magstep1
\font\twelvett=cmtt10 scaled \magstep1 \font\twelvebf=cmbx12
\font\twelverm=cmr12 \font\twelvei=cmmi12
\font\twelvess=cmss12 \font\twelvesy=cmsy10 scaled \magstep1
\font\twelvesl=cmsl12 \font\twelveex=cmex10 scaled \magstep1
\font\twelveit=cmti12
\font\tenss=cmss10
 
 \font\ninebf=cmbx7 scaled \magstep1
\font\ninerm=cmr7 scaled \magstep1 \font\ninei=cmmi7 scaled \magstep1
\font\ninesy=cmsy7 scaled \magstep1 
\font\eightrm=cmr7 scaled 1140 
 
\font\sevenbf=cmbx7 \font\sevenrm=cmr7 \font\seveni=cmmi7
\font\sevensy=cmsy7 

\catcode`@=11
\newskip\ttglue
\newfam\ssfam

\def\fourteenpoint{\def\rm{\fam0\fourteenrm}
\textfont0=\fourteenrm \scriptfont0=\tenrm \scriptscriptfont0=\sevenrm
\textfont1=\fourteeni \scriptfont1=\teni \scriptscriptfont1=\seveni
\textfont2=\fourteensy \scriptfont2=\tensy \scriptscriptfont2=\sevensy
\textfont3=\fourteenex \scriptfont3=\fourteenex \scriptscriptfont3=\fourteenex
\def\it{\fam\itfam\fourteenit} \textfont\itfam=\fourteenit
\def\sl{\fam\slfam\fourteensl} \textfont\slfam=\fourteensl
\def\bf{\fam\bffam\fourteenbf} \textfont\bffam=\fourteenbf
\scriptfont\bffam=\tenbf \scriptscriptfont\bffam=\sevenbf
\def\tt{\fam\ttfam\fourteentt} \textfont\ttfam=\fourteentt
\def\ss{\fam\ssfam\fourteenss} \textfont\ssfam=\fourteenss
\tt \ttglue=.5em plus .25em minus .15em
\normalbaselineskip=16pt
\abovedisplayskip=16pt plus 4pt minus 12pt
\belowdisplayskip=16pt plus 4pt minus 12pt
\abovedisplayshortskip=0pt plus 4pt
\belowdisplayshortskip=9pt plus 4pt minus 6pt
\parskip=5pt plus 1.5pt
\setbox\strutbox=\hbox{\vrule height12pt depth5pt width0pt}
\let\sc=\tenrm
\let\big=\fourteenbig \normalbaselines\rm}
\def\fourteenbig#1{{\hbox{$\left#1\vbox to12pt{}\right.\n@space$}}}

\def\twelvepoint{\def\rm{\fam0\twelverm}
\textfont0=\twelverm \scriptfont0=\ninerm \scriptscriptfont0=\sevenrm
\textfont1=\twelvei \scriptfont1=\ninei \scriptscriptfont1=\seveni
\textfont2=\twelvesy \scriptfont2=\ninesy \scriptscriptfont2=\sevensy
\textfont3=\twelveex \scriptfont3=\twelveex \scriptscriptfont3=\twelveex
\def\it{\fam\itfam\twelveit} \textfont\itfam=\twelveit
\def\sl{\fam\slfam\twelvesl} \textfont\slfam=\twelvesl
\def\bf{\fam\bffam\twelvebf} \textfont\bffam=\twelvebf
\scriptfont\bffam=\ninebf \scriptscriptfont\bffam=\sevenbf
\def\tt{\fam\ttfam\twelvett} \textfont\ttfam=\twelvett
\def\ss{\fam\ssfam\twelvess} \textfont\ssfam=\twelvess
\tt \ttglue=.5em plus .25em minus .15em
\normalbaselineskip=14pt
\abovedisplayskip=14pt plus 3pt minus 10pt
\belowdisplayskip=14pt plus 3pt minus 10pt
\abovedisplayshortskip=0pt plus 3pt
\belowdisplayshortskip=8pt plus 3pt minus 5pt
\parskip=3pt plus 1.5pt
\setbox\strutbox=\hbox{\vrule height10pt depth4pt width0pt}
\let\sc=\ninerm
\let\big=\twelvebig \normalbaselines\rm}
\def\twelvebig#1{{\hbox{$\left#1\vbox to10pt{}\right.\n@space$}}}

\def\tenpoint{\def\rm{\fam0\tenrm}
\textfont0=\tenrm \scriptfont0=\sevenrm \scriptscriptfont0=\fiverm
\textfont1=\teni \scriptfont1=\seveni \scriptscriptfont1=\fivei
\textfont2=\tensy \scriptfont2=\sevensy \scriptscriptfont2=\fivesy
\textfont3=\tenex \scriptfont3=\tenex \scriptscriptfont3=\tenex
\def\it{\fam\itfam\tenit} \textfont\itfam=\tenit
\def\sl{\fam\slfam\tensl} \textfont\slfam=\tensl
\def\bf{\fam\bffam\tenbf} \textfont\bffam=\tenbf
\scriptfont\bffam=\sevenbf \scriptscriptfont\bffam=\fivebf
\def\tt{\fam\ttfam\tentt} \textfont\ttfam=\tentt
\def\ss{\fam\ssfam\tenss} \textfont\ssfam=\tenss
\tt \ttglue=.5em plus .25em minus .15em
\normalbaselineskip=12pt
\abovedisplayskip=12pt plus 3pt minus 9pt
\belowdisplayskip=12pt plus 3pt minus 9pt
\abovedisplayshortskip=0pt plus 3pt
\belowdisplayshortskip=7pt plus 3pt minus 4pt
\parskip=0.0pt plus 1.0pt
\setbox\strutbox=\hbox{\vrule height8.5pt depth3.5pt width0pt}
\let\sc=\eightrm
\let\big=\tenbig \normalbaselines\rm}
\def\tenbig#1{{\hbox{$\left#1\vbox to8.5pt{}\right.\n@space$}}}
\let\rawfootnote=\footnote \def\footnote#1#2{{\rm\parskip=0pt\rawfootnote{#1}
{#2\hfill\vrule height 0pt depth 6pt width 0pt}}}

\def\tenfoot{\tenpoint\hskip-\parindent\hskip-.1cm}

\overfullrule=0pt
\twelvepoint
\oneandahalfspace
\def\sbullet{\raise.2em\hbox{$\scriptscriptstyle\bullet$}}
\nofirstpagenotwelve
\hsize=16.5 truecm
\baselineskip 15pt

\def\ft#1#2{{\textstyle{{#1}\over{#2}}}}

\oneandahalfspace
\rightline{CTP TAMU--27/93}
\rightline{IC/93/165}
\rightline{hep-th/9307043}
\rightline{July 1993}

\vskip 2truecm
\centerline{\bf BRST Operator for Superconformal Algebras }
\centerline{\bf with Quadratic Nonlinearity}
\vskip 1.5truecm
\centerline{Z. Khviengia and E. Sezgin\footnote{$^\dagger$}{\tenfoot
\sl  Supported in part by the National Science Foundation, under grant
PHY-9106593.}}

\vskip 1.5truecm
\centerline{\it  Center
for Theoretical Physics,
Texas A\&M University,}
\centerline{\it College Station, TX 77843--4242, USA.}
\vskip 1.5truecm
\AB\singlespace
   We construct the quantum BRST operators for a large class of superconformal
and quasi--superconformal algebras with quadratic nonlinearity. The only
free parameter in these algebras is the level of the (super) Kac-Moody
sector. The nilpotency of the quantum BRST operator imposes a condition on
the level. We find this condition for (quasi) superconformal algebras with
a Kac-Moody sector based on a simple Lie algebra and for the $Z_2\times
Z_2$--graded superconformal algebras with a Kac-Mody sector based on the
superalgebra $osp(N\vert 2M)$ or $s\ell(N+2\vert N)$.

\AE\oneandahalfspace

\np

\noindent
{\bf 1. Introduction}
\bigskip
Superconformal algebras in two dimensions play an important role in superstring
theories.
With certain assumptions, such as linearity, the classification of
$so_N$--extended algebras  was given long ago [1].
They contain a single spin 2 generator and multiplets of other generators of
spin decreasing by half  units down to minimum spin $2-{1\over 2}N$. Thus, for
$N>4$ there are negative  dimension generators which are problematic in any
application to string theory. In fact, so far only $N=1,...,4$
superconformal algebras have found  applications in string theory.

It is clearly of interest to examine the possibility of having higher extended
superconformal algebras which can be used in constructing new types of
superstring theories. Relaxing the requirement of linearity, higher
extended superconformal algebras do actually exist. In [2] and [3], certain
types of
such algebras have been found. They contain the energy momentum tensor $T$, a
multiplet of spin 3/2 currents $G_i$, in the fundamental representation of
$so_N$
or $u_N$, and a multiplet of spin 1 currents $J^a$, in the adjoint
represenation
of the same algebras. A characteristic feature of these algebras is that the
OPE
of two spin 3/2 currents contains an operator bilinear in spin 1 currents.
This is very similar to the case of  $W_3$
algebra where the OPE of two spin 3 currents contains the square of the
energy-momentum tensor [4].

A convenient way of characterizing the quadratically nonlinear algebras is to
specify the pair $(g,\rho)$ where $g$ is the (super) Lie algebra and $\rho$ is
the representation carried by the spin 3/2 currents. Various possibilities
for this pair have been found [5-10] in addition to those in [2,3].
All the known cases are listed in Tables 1-3. In particular, in the case of
Table 1, as shown by Fradkin and Linetsky [5], the necessary and sufficient
condition for the algebra to exist is the following identity
$$
(\lambda^a_{ij}\lambda^a_{k\ell}-\lambda^a_{jk}\lambda^a_{i\ell})=
{\epsilon C_\rho\over d+\epsilon}
\big(\eta_{ij}\eta_{kl}+\eta_{jk}\eta_{i\ell}-2\eta_{ki}\eta_{j\ell}\big)\ ,
$$
where $\lambda^a_{ij}$ are te generators and $\eta_{ij}$ is the
invariant tensor of the algebra $g$, and $C_\rho$ is the
eigenvalue of the second
Casimir in the $d$-dimensional representation $\rho$, and $\epsilon=-1$ for
superconformal algebras, $\epsilon=1$ for quasiconformal algebras.

In the case of
$N$--extended superconformal algebra for $N=3,4$, the quadratic nonlinearity
can be removed by the introduction of spin 1/2 and spin 0 operators [11].
Therefore we shall focus our attention to the case $N>4$. It should also be
noted that the algebras $g$ can be taken to be complex. By imposing suitable
reality conditions one can then obtain their real forms.

The algebras listed in Table 3 are $Z_2 \times Z_2$ graded since the
Kac-Moody sector itself contains bosonic and fermionic generators. In the
limit of purely bosonic or fermionic Kac-Moody generators, these algebras
reduce to some of the (quasi) superconformal algebras listed in Tables 1
and 2.

The difference between superconformal and quasisuperconformal algebras is that
the spin 3/2 currents are fermionic in the former case and
bosonic in the latter case. In all these algebras the OPEs of the operators
$T$,
$G_i$ and $J^a$ are the standard ones with the exception of the OPE of $G_i$
with $G_j$ which will contain a $JJ$ term. The explicit form of these
OPE's will be exhibited in the following sections. Suffices to mention that
these algebras
admit nontrivial central extensions and that all the parameters occuring in
the
algebra including the  central extensions are determined in terms of the level
of the Kac-Moody sector of the algebra, which is the only free parameter.

An elegant general formula for the classical and
quantum BRST operator for a large class of quadratically nonlinear algebras
has been derived by Scoutens, Sevrin
and van Nieuwenhuizen [12]. For the $so_N$--extended
superconformal algebra it was found that the existence of the BRST operator
fixes the level to be $k=-2(N-3)$. Hence for the cases of interest, i.e.
$N>4$, the level is negative and consequently the algebra does not admit
unitary representations. In [12], it was furthemore found that a quantum
BRST operator does not exist for the case of $u_N$--extended
superconformal algebra. In this paper we shall generalize these results by
constructing the quantum BRST operator for all the cases listed in Table 1
(which includes the quasi--superconformal algebras) and $Z_2\times
Z_2$--graded algebras with a Kac-Moody sector based on the
superalgebra $osp(N\vert 2M)$ or $s\ell(N+2\vert N)$ (see Table 3).
We shall give the explicit form of the quantum BRST operators for these cases
and we shall find the conditions on the levels imposed by the nilpotency of
the BRST operator. For all the algebras listed in Table 1 and for
the $Z_2 \times Z_2$--graded algebra
based on $s\ell(N+2\vert N)$, we find that the level is still negative
(eqs. (2.8a) and (3.15), respectively). In the case of  $Z_2 \times
Z_2$--graded algebra based on $osp(N\vert 2M)$ the level can be negative or
positive (eq. (3.8a)).
The next two sections are devoted to the derivation of these results. Sec. 4
contains comments on the implications of these results for the existence of
unitary representations and on a number of other issues.
\bigskip
\noindent{\bf 2. BRST Operator for (Quasi) superconformal Algebras
With  Bosonic \hfil\break}
\noindent{\bf  Kac-Moody Sector}
\bigskip
The algebras of the type listed in Table 1  are
generated by the
energy-momentum tensor $T(z)$, the dimension 3/2 supercurrents $G^i(z),
i=1,...,{\rm dim}\ \rho := d$ and the dimension 1 currents $J^a(z),
a=1,...,{\rm dim}\ g := D$. The operator product algebra is as follows:
$$
\eqalignno{
T(z) T(\omega) =& { \ft12 c\over (z-\omega)^4}+ {2T(\omega)\over
(z-\omega)^2}+{\partial T(\omega)\over (z-\omega)} +\cdots \ , &(2.1a)\cr
T(z)G^i(\omega) =&{\ft32 G^i(\omega)\over (z-\omega)^2}+{\partial
G^i(\omega)\over (z-\omega)}+\cdots\ , &(2.1b)\cr
T(z)J^a(\omega) =&{J^a(\omega)\over (z-\omega)^2}+{\partial J^a(\omega)\over
(z-\omega)}+\cdots\ , &(2.1c)\cr
G^i(z)G^j(\omega) =& {b\eta^{ij}\over (z-\omega)^3}+ {\sigma \lambda^{ij}_a
J^a(\omega)\over (z-\omega)^2} +{\ft12 \sigma \lambda^{ij}_a
\partial J^a(\omega)\over (z-\omega)} +{2\eta^{ij} T(\omega)\over (z-\omega)}
\cr
& +{\gamma P^{ij}_{ab} (J^a J^b)(\omega)\over (z-\omega)}+\cdots \ ,&(2.1d)\cr
J^a(z) G^i(\omega) =& {-\lambda^{ai}_{~~j} G^j(\omega)\over (z-\omega)}+\cdots
\ , &(2.1e)\cr
 J^a(z) J^b(\omega) =& {-\ft12 k\psi^2 \delta^{ab}\over (z-\omega)^2}+
{f^{ab}_{~~~c} J^c(\omega)\over (z-\omega)}+\cdots \ ,&(2.1f)\cr}
$$
where the generators $\lambda^a_{ij}$  and the structure constants
$f_{ab}^{~~c}$ satisfy the relations
\footnote{$\dagger$}{We use the
conventions of Ref. [5b].  It may be useful to note, however, that
there are misprints in eqs. (5), (10.a,b), (11) and (14b) of
Ref. [5b]).}
$$
\eqalign{
\lambda_{aik}\lambda^k_{bj}-
\lambda_{bik}\lambda^k_{aj}
 = f_{ab}^{~~c} \lambda_{ci}^{~~j} \ ,  \quad\quad
\lambda_{aij}\lambda_b^{ji}=-i_\rho \psi^2 \delta_{ab}  \ , \cr
(\lambda^a_{ij}\lambda^{a}_{k\ell}-\lambda^a_{jk}\lambda^{a}_{i\ell})=
{2\epsilon\over
\sigma_0} (\eta_{ij}\eta_{kl}+\eta_{jk}\eta_{i\ell}-2\eta_{ki}\eta_{j\ell}),
\qquad  \sigma_0=2(d+\epsilon)/C_\rho\ ,  \cr}\eqno(2.2)
$$
and the quadratic nonlinearity is defined by  $(JJ)(\omega) :=
{1\over 2\pi i}\oint d\zeta {J(\zeta)J(\omega)\over (\zeta-\omega)}$.
The Cartan-Killing metric $g^{ab}$ is defined as follows
$$
g_{ab}=f_{ac}^{~~d}f_{bd}^{~~c} =-C_v \delta_{ab}\ . \eqno(2.3)
$$
The Lie algebra is taken to be complex for the time being. The real forms
and their implications for constructing  unitary represantations will be
briefly disussed in Sec. 4.
The Dynkin index $i_\rho$ of the representation $\rho$ is defined by
$i_\rho={dC_\rho\over D\psi^2}$, where $C_\rho$ is the eigenvalue of the
second Casimir in the representation $\rho$ defined by  $
\lambda^a_{ik}\lambda_a^{kj} =-C_\rho \delta_i^j$ and $\psi^2$ is the square
of  the longest root. In a
convention where the shortest root squared is 2 for all the Lie algebras,
the value of $\psi^2$ is 2 for the simply laced Lie algebras (i.e. $A_n$,
$D_n$, $E_6$, $E_7$, $E_8$), 4 for $B_n$, $C_n$, $F_4$ and 6 for $G_2$.
The central extension in the Kac-Moody algebra is parametrized such
that the unitary highest weight representations exist for positive
integer values of $k$. $C_v$ is the
eigenvalue of the second Casimir in the adjoint representation of $g$ related
to the dual Coxeter number $h_g^\vee$ by $C_v=\psi^2 h_g^\vee$.
The raising and lowering of the indices $i,j,..$ is done by the
metric $\eta_{ij}=-\epsilon
\eta_{ji}$, satisfying the relation $\eta_{ik}\eta^{jk}=\delta_i^j$, by the
rule: $V^i=\eta^{ij}V_j$ and $V_i=V^j \eta_{ji}$, for any quantity $V$.
The parameter $\epsilon=-1$ for the superconformal algebras and $\epsilon=+1$
for the quasisuperconformal algebras. For $\epsilon=-1$ the currents
$G^i(z)$ are fermionic, while for   $\epsilon=+1$ they are bosonic.

The tensor $P^{ab}_{ij}$ is defined by
$$
P^{ab}_{ij}=  \lambda^a_{ik}\lambda^{bk}_{~~j}+
          \lambda^b_{ik}\lambda^{ak}_{~~j}+
{2\over \sigma_0} \eta_{ji}\delta^{ab}\ . \eqno(2.4)
$$
Note the symmetry properties: $P^{ab}_{ij}=-\epsilon P^{ab}_{ji}$ and
$\lambda^a_{ij}=\epsilon \lambda^a_{ji}$. Thus
$\eta^{ij}\lambda^a_{ij}=0$. The OPE algebra (2.1) closes
provided that the parameters occuring in the algebra obey the following
relations [5]
$$
\eqalignno{
\gamma =& {d(d+\epsilon)\over \psi^4 D i_\rho\eta },\qquad \eta :=
k+h_g^\vee+\epsilon i_\rho\ , &(2.5a) \cr
\sigma =& {2d\over \psi^2 D i_\rho\eta}[(d+\epsilon)(k+h_g^\vee)-Di_\rho] \ ,
&(2.5b) \cr
b=& {k\psi^2\sigma\over 2}\ , &(2.5c)\cr
c =&\ft32 b +{k\over \eta}(D+\epsilon d +1) \ . &(2.5d)\cr}
$$

We now turn to the construction of the BRST operator corresponding to the
above algebra. We introduce the pairs of ghosts $(b,c)$, $(\beta^i,\gamma_i)$
and $(r^a, s_a)$, corresponding to the generators $T, G^i$ and $J^a$,
respectively. The ghosts $(c,\gamma^i, s^a)$ have ghost number $1$ and
conformal dimension $(-1,-\ft12, 0)$, respectively, while the antighosts
$(b,\beta_i,r_a)$ have ghost number $-1$ and conformal dimension $(2,\ft32,
1)$, respectively. They satisfy the following OPEs
$$
c(z)b(\omega)={1\over (z-\omega)}+\cdots,\qquad
\gamma^i(z)\beta_j(\omega)={\delta^i_j\over (z-\omega)}+\cdots,\qquad
s^a(z)r^b(\omega)={\delta^{ab}\over (z-\omega)}+\cdots \ . \eqno(2.6)
$$

Using the result of
[12] we make an ansatz for the BRST operator depending on a number of
parameters. We then verify fully the nilpotency of the BRST operator, which
fixes all these parameters and in addition  imposes conditions on
the parameters of the algebra (2.1).  For the BRST operator we find the
following result:
$$
\eqalign{
  Q=& cT+\gamma^iG_i+s^a J_a+bc\partial c+\beta_i\big(\ft12 \gamma^i\partial
c-\partial\gamma^i c\big)\cr
 &-r_as^a\partial c-b\gamma_i\gamma^i +\lambda_{ai}^{~~j}\big( -\ft12
\epsilon\sigma_0 r^a\gamma^i\partial\gamma_j+s^a\beta^i\gamma_j\big)
-\ft12 f_{ab}^{~~c} r_cs^as^b\cr
& -\ft12 \gamma P^{ab}_{ij}J_ar_b\gamma^i\gamma^j-\ft1{24}\gamma^2 P^{ab}_{ij}
P^{cd}_{kl}f_{ac}^{~~e}
r_b r_d r_e\gamma^i\gamma^j\gamma^k\gamma^\ell \ .\cr}\eqno(2.7)
$$
Using the relations (2.5a,b) and various group theoretical relations provided
in [5], we find that the above BRST operator is nilpotent provided that
the cental extensions satify the following relations:
$$
\eqalignno{
k=& -2(h_g^\vee+\epsilon i_\rho)\, &(2.8a)\cr
c=& 26+11\epsilon d+2D\ , &(2.8b)\cr
b=& 16+6\epsilon d \ . &(2.8c)\cr}
$$
The first condition comes from cancellation requirement of terms bilinear in
$s^a$. Similarly the second condition comes from the
cancellation of the terms bilear in $c$, and the last
condition  from the cancellation of the terms bilinear in  $\gamma^i$.  In
(2.8b) the central charge equals the sum of contributions
$2(-1)^{2s}(6s^2-6s+1)$ from each generator of conformal dimension $s$ with  an
aditional factor of $-\epsilon$ for spin 3/2 generator. The relations
(2.8b,c) agree with (2.5d,c),
upon the use of (2.8a) and particular values of various group theoretical
quantities
listed in Table 1. The most crucial new information
implied by the existence of
the quantum BRST operator is the condition  (2.8a) on the Kac-Moody level $k$.
{}From Table 1 we see that $k$ will always be negative
(in the case of first entry in Table 1, for $N>4$), which means difficulty in
obtaining unitary representations of the  algebra, at least in the case
when its compact form is considered. In particular, we see that
taking the spin 3/2 currents to be bosonic does not change this
conclusion.
However, considering the noncompact form of the algebras
listed in Table 1 may have interesting consequences for the existence of
unitary representations. This point will be discussed further in Sec. 4.

We close this section with a comment on the structure of the BRST
operator (2.7). The form of all the cubic in ghost terms follows from the
structure constants of the linear part of the algebra. The quartic in
ghost term is suggested by treating the $PJJ$ term in the algebra
as `linear' in $J$ with $PJ$ as a field dependent structure `constant'.
The occurence of the last term which is seventh order in ghost field,
which is needed for the nilpotency of the BRST operator, can be
understood from a more systematic point of view from the work of Ref. [12].

 \bigskip
\noindent{\bf 3. BRST Operator for Quadratically Nonlinear Superconformal
\hfil\break}
\noindent{\bf Algebras With  Super Kac-Moody Sector}
\bigskip
\centerline{\it The Case of $osp(N\vert 2M)$ }
\bigskip
In this case the dimension 3/2 current $G^i(z), i=1,...,N+2M$ carries the
fundamental representation of the superalgebra $osp(N \vert 2M)$, and the
spin 1
currents $J^a(z), a=1,...,dim\ g$ are in the adjoint representation of the
same
superalgebra. The first three OPE relations in (2.1) remain the same in
present case, while the remaining ones read
$$
\eqalignno{
G^i(z)G^j(\omega) =& {b\eta^{ji}\over (z-\omega)^3}+ {\sigma\lambda^{ij}_a
J^a(\omega)\over (z-\omega)^2} +{\ft12 \sigma \lambda^{ij}_a
\partial J^a(\omega)\over (z-\omega)} -{2\eta^{ji} T(\omega)\over (z-\omega)}
\cr
& +{\gamma P^{ij}_{ab} (J^a J^b)(\omega)\over (z-\omega)}+\cdots \ ,&(3.1b)\cr
J^a(z) G^i(\omega) =& {(-1)^{a+1}\lambda^{ai}_{~~j} G^j(\omega)\over
(z-\omega)}+\cdots \ , &(3.1b)\cr
 J^a(z) J^b(\omega) =& {-k \eta^{ba}\over (z-\omega)^2}+
{f^{ab}_{~~~c} J^c(\omega)\over (z-\omega)}+\cdots \ .&(3.1c)\cr}
$$
The Grassmann parity of the indices $a,i$ is indicated by the same letters
and $i=1$ for $so_N$, and $i=0$ for $sp_{2M}$, while $a=0$ for the
bosonic generators and $a=1$ for the fermionic generators.
The boson-fermion parity of the currents, denoted by $n$ are:
$n(T)=0,\ n(G^i)=i,\ n(J^a)=a$. Thus, $G^i$ is  fermionic for $i=1,...,N$ and
bosonic for $i=N+1,...,N+2M$. The metric $\eta_{ab}$ is defined in terms of
the structure constans as follows
$$
f^{cda}f^b_{~dc}=-2(d_s -2)\eta^{ab}\ , \qquad d_s := N-2M\ .\eqno(3.2)
$$
The symmetry properties of the metrics are: $\eta_{ab}=(-1)^{ab}
\eta_{ba}\ ,\eta_{ij}=(-1)^{ij+1}\eta_{ji}$. Note also that
$\eta_{ac}\eta^{bc}=\delta_a^b$ and $\eta^{ik}\eta_{jk}=\delta_i^j$.
The $osp(N\vert 2M)$ generators $\lambda^a_{ij}$ obey the algebra
$$
\lambda_{aik}\lambda^k_{bj}+(-1)^{ab+1}\lambda_{bik}\lambda^k_{aj}
=-f_{bac}\lambda^c_{ij}\ .  \eqno(3.3)
$$
The structure constants are totally graded antisymmetric, e.g.
$f_{abc}=(-1)^{ab+1}f_{bac}$. The generators satisfy the crucial identity
$$
\lambda^a_{ij}\lambda_{ak\ell}=-\eta_{kj}\eta_{\ell i}
+(-1)^{ij+1}\eta_{ki}\eta_{\ell j} \ . \eqno(3.4)
$$
This identity agrees with (2.2) for the case of
$so_N$, but differs by an overall minus sign for the case of $sp_{2M}$.
The reason for this difference is that the above equation
involves  fermionic as well as bosonic components, and it is such that the
symmetries of the left hand are  maintained by the right hand side.

The tensor $P^{ab}_{ij}$ is now defined by
$$
P^{ab}_{ij}=\lambda^a_{ik}\lambda^{bk}_{~~j}
+(-1)^{ij+1}\lambda^a_{jk}\lambda^{bk}_{~~i}+
2 \eta^{ab}\eta_{ji}\ . \eqno(3.5)
$$
Note that $\lambda^a_{ij}=(-1)^{ij}\lambda^a_{ji}$ and the Grassmann
parities obey the relation $a+i+j=0$, while
$P^{ab}_{ij}=(-1)^{ba}P^{ba}_{ij}=(-1)^{ij+1}P^{ab}_{ji}$
and the Grassmann parities satisfy the relation: $a+b+i+j=0$.

   The OPE algebra (2.1a-c) and (3.1) closes provided that [10]
$$
\eqalign{
\gamma=& {1\over 2(k+d_s-3)}\ , \cr
\sigma=& 2\gamma(2k+d_s-4)\ , \cr
b=&k\sigma\ ,\cr
c=& k\gamma(6k+d_s^2-10)\ .   \cr}\eqno(3.6)
$$
In fact, the above equations can be obtained from those in (2.5) by
evaluating them for $so_N$ and then making the substitution:  $d\rightarrow
d_s$. Thus they
clearly agree with the $so_N$ case ($M=0$). The comparison with the
$sp_{2M}$ case ($N=0$) is more subtle due to the difference
of normalizations in (3.4), (3.1c) and their counterparts for $sp_{2M}$.

The construction of  the quantum BRST charge proceeds as before.
We introduce the ghost pairs $(b,c)$, $(\beta^i, \gamma_i)$ and $(r^a, s_a)$
which obey the OPEs given in (2.6). We then make an ansatz similar to that
in (2.7) and by an explicit calculation determine all the coefficients.
Extreme care has to be exercised in dealing with the Grassmann
parities. At the end we find the following result:
$$
\eqalign{
  Q=& cT+\gamma^iG_i+s^a J_a+bc\partial c+\beta_i(\ft12 \gamma^i\partial
c-\partial\gamma^i c)\cr
 &-r_as^a\partial c-b\gamma^i\gamma_i
-r_a\lambda_i^{a~j}\gamma_j\partial\gamma^i
+s_a \lambda^{aij}\beta_j\gamma_i
+\ft12 f_b^{~ac}r_cs_as^b \cr
& -\ft12 \gamma (P^{ab})_i^{~j}r_aJ_b\gamma_j\gamma^i
-\ft1{24}\gamma^2
(P^{ab})_i^{~j}(P^{cd})_k^{~\ell}f_{aec}r^er_dr_b\gamma_j\gamma^i
\gamma_\ell\gamma^k\ .  \cr}\eqno(3.7)
$$
This BRST operator is
nilpotent provided that the folowing relations hold:
$$
\eqalignno{
k=& -2 (d_s-3)\ , &(3.8a) \cr
c=& 26-12d_s+d_s^2\ ,&(3.8b)\cr
b=& 16-6 d_s \ . &(3.8c)\cr}
$$
Note that $sdim\ g= \ft12 N(N-1)+M(2M+1)-2MN=d_s(d_s-1)$, where we
recall that the superdimension $d_s$ is defined by $d_s=N-2M$.
We have also used (3.2) and the relation
$\lambda^a_{ij}\lambda^{bji}=-2 \eta^{ab}$. The above conditions agree for the
special case of $so_N$ with
$M=0$, while the comparison for the case of $sp_{2M}$ is more subtle as
mentioned earlier, due to the different normalizations chosen
in (3.4), (3.1c) and in their counterparts for $sp_{2M}$.
Note that now the level $k$ is a positive integer for $N<2M+3$.

The quantum BRST operator for the last case in Table 3 proceeds exactly in
the same manner as described above, and its further discussion will
be omitted here. Finally we discuss the $Z_2\times Z_2$ graded algebra
based on $s\ell(N+2\vert N)$.
\bigskip
\centerline{\it The case of $s\ell(N+2\vert N)$}
\bigskip
First let us discuss briefly the case of $g\ell(N\vert M)$. A real
form of this algebra based on the compact superalgebra $u(N|M)$
was constructed in [10]. The limit $M=0$ coincides with the well known
case based on $su(N)\times u(1)$ [2,3]. Interestingly enough, in [12] it
was found that the quantum BRST operator does not exist for this
case, because a condition that relates the Kac-Moody levels of $u(N)$ and
$u(1)$ contradicts the condition implied by the closure of the OPE algebra.
Given the results of the previous section, we expect that a similar
situation will occur in the case of $g\ell(N\vert M)$, with $N$ of [12] being
replaced by the superdimension $d_s=N-M$. Thus, we expect that the quantum
BRST operator does not exist for the $g\ell (N\vert M)$ case. Nonetheless,
an interesting situation arises for $d_s=2$. As was shown in [10],
in that case the $g\ell(1)$ current decouples completely. Then we have
$s\ell (N+2\vert N)$ for which an invertible Cartan-Killing metric
$g_{ab}$ defined as in (2.3) does exist. An interesting property of
this algebra is that, despite the fact that it looks like the usual $N=4$
superconformal algebra with $SU(2)$ Kac-Moody sector, it is however
different in that the quadratic nonlinearity is still there, and
there are $2(N+2)$ anticommuting and $2N$ commuting
spin 3/2 generators. Below, we shall consider this
case, for which, as we shall see there does exist a quantum BRST operator.

For $s\ell(N+2\vert N)$ case, the spin 3/2 currents are $G^i(z)$
and $G_i(z)$, $ i=1,...,2N+2$. These are independent, as there is no
metric to raise and lower indices. The boson-fermion parities are
$n(G^i)=n(G_i)=i$. As mentioned above there are $2(N+2)$
fermionic and $2N$ bosonic spin 3/2 generators. The nontrivial OPE's in
this case are [10]
$$
\eqalign{
G_i(z)G^j(\omega) =& {8(-1)^i\delta_i^j \over (z-\omega)^3}+
{2(-1)^k\lambda_i^{a~j} J_a(\omega)\over (z-\omega)^2} +
   {(-1)^k\lambda_i^{a~j} \partial J_a(\omega) \over (z-\omega)}
+{2(-1)^i\delta_i^j T(\omega)\over (z-\omega)} \cr
& {\gamma (P^{ab})_i^j (J_a J_b)(\omega)\over (z-\omega)}+
\cdots \ ,\cr
J_a(z) G_i(\omega) =& -{\lambda_{ai}^{~~j} G_j(\omega)\over
(z-\omega)}+\cdots \ , \cr
J_a(z) G^i(\omega) =&
{(-1)^{i+ij}G^j(\omega) \lambda_{aj}^{~~i}\over (z-\omega)}+\cdots \ , \cr
J^a(z) J^b(\omega) =& {-k \eta^{ba}\over (z-\omega)^2}+
{f^{ab}_{~~~c} J^c(\omega)\over (z-\omega)}+\cdots \ ,
\cr} \eqno(3.9)
$$
where
$$
(P^{ab})_i^{~j}= \lambda_i^{ak}\lambda_k^{bj}+(-1)^{ab}\lambda_i^{b~k}
\lambda_k^{a~j} +2\delta_i^j \eta^{ab}\ . \eqno(3.10)
$$
The parameter $\gamma$ and the Virasoro central extension $c$ are
related to the level $k$ as follows [10]
$$
    \gamma=-{1\over 2(k+2)}\ ,\quad\quad c=3k\ . \eqno(3.11)
$$
The generators are supertraceless,
i.e $(-1)^{i+1} \lambda_i^{ai}=0$ and obey the graded commutation rule:
$\lambda_i^{ak}\lambda_k^{bj}+(-1)^{ab+1}\lambda_i^{bk}\lambda_k^{aj}
=-f^{ba}_{~~c}\lambda_i^{c~j}$. Further important identities are
$$
\eqalign{
&\lambda_i^{aj}\lambda_{ak}^{~~\ell}=\delta_i^j\delta_k^\ell+
2(-1)^k\delta_i^\ell\delta_k^j\ , \cr
&f^{cda}f^b_{~dc}=-8\eta^{ab}\ , \qquad (-1)^{i+1}\lambda_i^{aj}
\lambda_j^{bi}=-2 \eta^{ab}\ .\cr}  \eqno(3.13)
$$

In order to construct the BRST operator, we now introduce the ghost pairs
$(b,c)$, $(\beta_i, \gamma^i)$, $(\beta^i, \gamma_i)$ and $(r_a, s_a)$. By
explicit calculation we then find the following result for the BRST operator:
$$
\eqalign{
  Q=& cT+\gamma^iG_i+\gamma_i G^i+s^a J_a+bc\partial c+
\beta_i(\ft12 \gamma^i\partial c-\partial\gamma^i c)+
\beta^i(\ft12 \gamma_i\partial c-\partial\gamma_i c)+\cr
& -r_as^a\partial
c+2b\gamma_i\gamma^i +\lambda_{ai}^{~~j}r^a\big( \gamma_j\partial\gamma^i
-\partial\gamma_j\gamma^i)\big) \cr
&+\lambda_i^{aj}s_a\big(\gamma_j\beta^i-\beta_j\gamma^i\big)
+\ft12 f_b^{~ac}r_cs_as^b \cr
&+\ft14 (P_{ab})_i^{~j}r^aJ^b\gamma_j\gamma^i
+\ft1{24}
(P_{ab})_i^{~j}(P_{cd})_k^{~\ell}f_{dae}r^c r_e r^b\gamma_j\gamma^i
\gamma_\ell\gamma^k\ .  \cr}\eqno(3.14)
$$
The nilpotency of this operator imposes the restriction
$$
k=-4\ ,  \eqno(3.15)
$$
which in particular implies that $c=-12$. Unitary highest weight
representations of the algebra presumably do not exist for this value of the
level (see below).
\bigskip
\noindent{\bf 4. Comments}
\bigskip
So far we have considered complex Lie (super) algebras. The real forms of the
algebras listed in Table 1 are well known, and the real form of the
superalgebras
can be found in [13]. Care must be exercised in extending these reality
conditions to the full affine algebras considered here. In [5] it is
argued that in the case of superconformal algebras $(\epsilon=-1)$, real
forms exist with $g$ and $\rho$ real, while this is not possible in the case
of quasiconformal algebras with the exception of $s\ell(N+2,C)$ giving
rise to the real form $su (N+1,1)$ [9]. An interesting possibility
when  noncompact real forms exist is that, while the BRST
condition on the level is the one that forbids unitary highest weight
representations in the compact case, one may utilize a coset construction
upon which one imposes $H$ invariance condition on the acceptable states,
where $H$ is the maximal compact subgroup of the noncompact group. This
eliminates the negative norm states, making it possible to construct
unitary representations possibly with acceptable conditions on
the level. This phenomenon has been  studied for a number of cases [14]. (In
particular, see Ref. [15] for an interesting use of $SO(d-1,2)$ Kac-Moody
algebra to built a string theory in anti de Sitter space).
 Of course one would have to verify that the combined system of Virasoro
plus Kac-Moody algebras, taking into account the quadratic nonlinearities
as well, posesses a unitary represenation. If a
super Sugawara construction exists, the unitarity of the Kac-Moody
sector will ensure the unitarity of the whole system.

As far as the existence of the unitary highest weight representations of
Kac-Moody superalgebras are concerned, this problem has been addressed
in [16]. It was found that among the compact form of all such
algebras only $su(N+1\vert 1)$ and $osp(2\vert 2M)$ admit unitary highest
weight representations, with a suitable condition imposed on the level $k$.
They also computed the values of the Virasoro central charges for the
existence of super Sugawara construction [17,16]. Among the compact real
forms of the superalgebras considered here which may arise, only
$su(3\vert 1)$ and $osp(2\vert 2M)$
admit unitary highest weight representations, with suitable
restrictions
on the level $k$ [16], which however are in conflict with the restrictions
found
here from the existence of the BRST operator. As to the noncompact
real forms of the super Kac-Moody algebras considered here which may
arise, only those based on  $osp(N\vert 2,\ R)$ admit unitary highest weight
representations, again with suitable restrictions on the level $k$ [16]. Some
of those cases appear not to be in conflict with the restrictions found
here. Various aspects of this problem clearly deserves further investigation.

The problem of finding unitary representations of the quadratically
nonlinear algebras with the level conditions imposed by the existence of
the BRST operator is one of the important problems. A second
important problem is to find a spacetime realization of superconformal field
theories based on these algebras.
Some realizations of these algebras are known [18], but it is not clear how
they
can lead to a spacetime interpretation. In these realizations, group manifolds
seem naturally to arise but not Minkowskian spacetimes. It would be
interesting to see if
in a coset
construction of the type mentioned above, or possibly in a theory based on
a Wigner-In\"on\"u type contractions of the  algebras considered here,
a more promising geometrical framework lending itself to some
spacetime interpretation might emerge. It is certainly
worthwhile to investigate possible uses of these algebras in building the
internal sector of a novel string theory in the style of Gepner [19],
where $N=2$ superconformal theories are used in the construction of $N=1$
supersymmetric string theories in four dimensions. It would
also be interesting to see if these algebras could be used in certain
statistical systems where even nonunitary representation can have physical
interpretations.

Finally, it should be noted that the quadratically nonlinear algebras
considered here represent
a mildly nonlinear extension of the usual linear
superconformal algebras.
There are other, far more nonlinear versions  which allow  extended
supersymmetry beyond $N=4$. An interesting example is the $N=8$
superconformal algebra of
Ref. [20], based on a loop algebra of paralell transformations on seven
spheres. In [21], the emergence of this algebra in a twistor
formulation of the Green-Schwarz superstring is shown and a BRST operator is
constructed. Another example is provided by the $N=8$ conformal supergravity
theory of Ref. [22]. This theory has not been investigated in any detail so
far.

\bigskip\bigskip\bigskip
\centerline{\bf Acknowledgements}
\bigskip
We thank A.O. Barut, E. Bergshoeff, N. Berkovits, V. Dobrev, H. Lu, C.N. Pope
and S. Schrans for
helpful discussions. We also thank the Internatinal Center for Theoretical
Physics, where part of this work was done, for hospitality.

\np
\bigskip
\vbox{\tabskip=0pt\offinterlineskip
\def\tablerule{\noalign{\hrule}}
\halign to 300pt{\strut#&\vrule#\tabskip=0em plus2.5em&
\hfil#&\vrule#&\hfil#&\vrule#&\hfil#&\vrule#&
\hfil#&\vrule#&\hfil#&\vrule#
\tabskip=0pt\cr\tablerule
&&\omit\hidewidth $g$ \hidewidth
&&\omit\hidewidth $\rho$ \hidewidth
&&\omit\hidewidth $h_g^\vee$ \hidewidth
&&\omit\hidewidth $i_\rho$\hidewidth
&&\omit\hidewidth References \hidewidth &\cr\tablerule
&&\multispan9\hfil superconformal\quad ($\epsilon=-1$)\hfil&\cr\tablerule
&&$so_N$&&$N$&&$N-2$ &&1&& \quad [2,3]\qquad &\cr\tablerule
&&$so_7$&&$8_s$&&5 &&1&& \quad [6,7]\qquad &\cr\tablerule
&&$G_2$&&7&&4 &&1&&\quad  [6,7]\qquad &\cr\tablerule
&&\multispan9\hfil quasisuperconformal\quad
($\epsilon=1$)\hfil&\cr\tablerule
&&$sp_{2M}$&&$2M$&&$M+1$ &&$1/2$&&\quad  [8]\qquad &\cr\tablerule
&&$s\ell_6$&&20&&6 &&3&& \quad [5]\qquad &\cr\tablerule
&&$so_{12}$&&32&&10 &&4&& \quad [5]\qquad &\cr\tablerule
&&$E_7$&&56&&18 &&6&& \quad [5]\qquad &\cr\tablerule
&&$sp_6$&&14&&4 &&$5/2$&& \quad [5]\qquad &\cr\tablerule
&&$s\ell_2$&&4&&2 &&$5/2$&& \quad [5]\qquad &\cr\tablerule}}
\bigskip
\noindent{ Table 1.\ (Quasi) superconformal and algebras
with simple $g$ and irreducible $\rho$.}
\bigskip
\bigskip
\bigskip
\vbox{\tabskip=0pt\offinterlineskip
\def\tablerule{\noalign{\hrule}}
\halign to 300pt{\strut#&\vrule#\tabskip=0em plus2.5em&
\hfil#&\vrule#&\hfil#&\vrule#&\hfil#&\vrule#
\tabskip=0pt\cr\tablerule
&&\omit\hidewidth $g$ \hidewidth
&&\omit\hidewidth $\rho$ \hidewidth
&&\omit\hidewidth References \hidewidth &\cr\tablerule
&&\multispan5\hfil superconformal\quad ($\epsilon=-1$)\hfil&\cr\tablerule
&&$g\ell_N$&&$N\oplus {\bar N}$&&[2,3]\qquad &\cr\tablerule
&&$sp_2\oplus sp_{2M}$&&(2,2M)&&[5,6]\qquad &\cr\tablerule
&&\multispan5\hfil quasisuperconformal\quad ($\epsilon=1$)\hfil&\cr\tablerule
&&$g\ell_N$&&$N\oplus {\bar N}$&&[8,9]\qquad &\cr\tablerule
&&$sp_2\oplus so_N$&&(2,N)&&[5]\qquad &\cr\tablerule}}
\bigskip
\noindent{ Table 2.\ (Quasi) superconformal algebras with non-simple $g$
and/or reducible $\rho$.}
\bigskip
\bigskip
\bigskip

\vbox{\tabskip=0pt\offinterlineskip
\def\tablerule{\noalign{\hrule}}
\halign to 300pt{\strut#&\vrule#\tabskip=0em plus2.5em&
\hfil#&\vrule#&\hfil#&\vrule#&\hfil#&\vrule#
\tabskip=0pt\cr\tablerule
&&\omit\hidewidth $g$ \hidewidth
&&\omit\hidewidth $\rho$ \hidewidth
&&\omit\hidewidth References \hidewidth &\cr\tablerule
&&$osp(N\vert 2M)$&&$ N\vert 2M$ &&[10]\qquad &\cr\tablerule
&&$g\ell(N\vert M)$&&$N\vert M$&&[10]\qquad &\cr\tablerule
&&$sp_2\oplus osp(N\vert 2M)$&&$(2, N\vert 2M)$ &&[5]\qquad &\cr\tablerule
}}
\bigskip
\noindent{Table 3.\ $Z_2\times Z_2$ graded superconformal algebras.}

\np
\centerline{\bf References}
\bigskip
\item{1.} M. Ademollo et al., Phys. Lett. {\bf B208} (1976) 447; Nucl. Phys.
{\bf B111} (1976) 77.
\item{2.} V.G. Knizhnik, Theor. Math. Phys. {\bf 66} (1986) 68.
\item{3.} M. Bershadsky, Phys. Lett. {\bf B174} (1986) 285.
\item{4.} A.B. Zamolodchikov, Theor. Math. Phys. {\bf 65} (1986) 1205.
\item{5.} E.S. Fradkin and V. Y. Linetsky, Phys. Lett. {\bf B282} (1992)
         352;{\bf B291} (1992) 71.
 \item{6.} P. Bowcock, Nucl. Phys. {\bf B381} (1992) 415.
\item{7.} E.S. Fradkin and V.Y. Linetsky, Phys. Lett. {\bf B275} (1992) 345.
\item{8.} L.J. Romans, Nucl. Phys. {\bf B357} (1991) 549.
\item{9.} F.A. Bais, T. Tjin and P. van Driel, Nucl. Phys. {\bf B357}
(1991) 632;
\item{} A.M. Polyakov, in ``Physics and Mathematics of Strings'' (World
Scientific, 1990);
\item{} M. Bershadsky, Commun. MAth. Phys. {\bf 139} (1991) 71.
\item{10.} F. Defever, W. Troost and Z. Hasiewicz, Phys. Lett. {\bf B273}
(1991) 51.
\item{11.} P. Goddard and A. Schwimmer, Phys. Lett. {\bf B214} (1988) 209.
\item{12.} K. Schoutens, A. Sevrin and P. van Nieuwenhuizen, Commun. Math.
           Phys. {\bf 124} (1989) 87.
\item{13.} M. Parker, J. Math. Phys. {\bf 21} (1980) 689.
\item{14.} L.J. Dixon, M.E. Peskin and J. Lykken, Nucl. Phys. {\bf B325}
           (1989) 329;
\item{} I. Bars, Nucl. Phys. {\bf B334} (1990) 125;
\item{} V. Dobrev and E. Sezgin, Int. J. Mod. Phys. {\bf A6} (1991) 4699.
\item{15.} E.S. Fradkin and V.Y. Linetsky, Phys. Lett. {\bf B261} (1991) 26.
\item{16.} P.D. Jarvis and R.B. Zhang, Phys. Lett. {\bf B215} (1988) 695;
           Nucl. Phys. {\bf B313} (1989) 205.
\item{17.} P. Goddard, D. Olive and G. Waterson, Commun. Math. Phys.
           {\bf 112} (1987) 591.
\item{18.} K. Schoutens, Nucl. Phys. {\bf B314} (1989) 519;
\item{} P. Matthieu, Phys. Lett. {\bf B218} (1989) 185;
\item{} K. Ito, J.O. Madsen and J.L. Petersen, preprint, NBI-HE-92-81
        (October 1992).
\item{19.} D. Gepner, Nucl. Phys. {\bf B296} (1987) 757; Phys. Lett.
           {\bf B199} (1987) 380.
\item{20.} F. Englert, A. Sevrin, W. Troost, A. van Proeyen
            and  P. Spindel, J. Math. Phys. {\bf 29} (1988) 281.
\item{21.} N. Berkovits, Nucl. Phys. {\bf B358} (1991) 169.
\item{22.} E. Bergshoeff, H. Nishino and E. Sezgin, Phys. Lett. {\bf B218}
           (1987) 167.

 \end